# Anomalous spin dynamics after dual optical excitation

Sergii Parchenko[1], Peter M. Oppeneer[2], Andreas Scherz[1]

[1]European XFEL, Holzkoppel 4, 22869 Schenefeld, Germany

[2]Department of Physics and Astronomy, Uppsala University, Box 516, SE-75120 Uppsala, Sweden

*Corresponding author: sergii.parchenko@xfel.eu

**Abstract.** Ultrashort optical pulses are a cornerstone for manipulating electronic and magnetic states in materials on a femtosecond timescale. Conventional models assume that optical excitation primarily modifies the occupation of the electron energy levels without long-lasting altering of the coupling of individual electrons in certain processes. Here, we demonstrate that optical excitation with two femtosecond pulses that come from different directions fundamentally transforms the electron dynamics in copper, affecting the efficiency of angular momentum transfer between electrons and the lattice. Using time-resolved magneto-optical Kerr effect measurements, we reveal a ~2.5× increase in spin imbalance decay time following inverse Faraday effect excitation under dual-pump conditions compared to single-pulse excitation. This observation challenges the prevailing paradigm of ultrafast light-matter interactions, showing that dual optical excitation can transiently modify electron dynamics beyond simple changes in the energy levels occupancy. Our findings open new avenues for controlling quantum states through a dual pump approach, with implications for ultrafast spintronics and the design of novel light-driven states.

**Introduction**

Ultrafast optical excitation has emerged as a powerful tool for controlling quantum states in materials, enabling precise manipulation of electronic, magnetic, and structural properties on femtosecond timescales [1,2]. The ability to initiate and probe rapid transitions between quantum states has opened new frontiers in fields such as ultrafast magnetism [3,4,5,6], spintronics [7,8,9], and quantum systems [10,11,12]. By leveraging femtosecond laser pulses, researchers can transiently alter material properties, such as magnetization and charge transport, providing insights into the fundamental interactions governing quantum systems. In magnetically ordered materials, ultrafast optical excitation has been shown to induce a variety of dynamic responses, including ultrafast demagnetization [13,14], magnetization precession [15], all-optical magnetization switching [16,17] and other ultrafast magnetic processes [18,19,20]. These phenomena are mediated by photon-electron energy transfer, where photoexcited electrons subsequently interact with the lattice and other electrons to drive the observed dynamics. According to the prevailing paradigm, optical excitation of electrons in metals acts primarily as a thermal stimulus - it elevates the electron temperature while leaving fundamental electronic properties unaffected. This framework makes two key assumptions: i) the excitation transiently heats the electron subsystem without altering intrinsic coupling strengths between electrons and the surrounding, and ii) all system parameters return to their equilibrium values following relaxation. Such thermalization assumptions form the basis of most ultrafast magnetism models, where light-matter interaction is treated exclusively as an electron heating mechanism that preserves the system's fundamental interaction parameters [1].

However, recent experiments using dual-pump optical excitation – where two ultrashort laser pulses incident from different directions with the same polarization are applied simultaneously – have challenged this conventional view. In magnetically ordered metallic materials, dual-pump excitation has been shown to significantly alter magnetization dynamics, including increased recovery times after demagnetization [21], reduced threshold fluences for all-optical switching [22], and extended lifetimes of magnetization precession [23]. These observations are in contrast to the expectation that dual-pump excitation would simply act as an "ultrafast electron heater" similar to single-pump action, suggesting instead that dual-pump excitation introduces a new factor that modifies the induced spin dynamics. One appealing hypothesis is that dual optical excitation alters the efficiency of angular momentum transfer by modifying the electron-electron and electron-lattice scattering. This would imply that certain properties of electrons themselves, rather than just the occupation of energy states, are transiently changed by the dual-pump excitation.

To explore this hypothesis, we carry out investigations on nonmagnetic copper, a material devoid of long-range magnetic order and collective spin interactions. By focusing on a nonmagnetic system, we can isolate the dynamics of individual electrons and probe how dual optical excitation affects their behavior. In this study, we use circularly polarized ultrashort laser pulses to induce spin dynamics in copper via the inverse Faraday effect (IFE), a process that generates a transient magnetic moment in response to optical excitation [24,25]. We compare the dynamics induced by single-pulse and dual-pump excitation, revealing a striking increase in the relaxation time of the induced magnetic moment under dual-pump conditions. This result suggests that dual optical excitation modifies the efficiency of angular momentum transfer between electrons and lattice, challenging the assumption that electron coupling with the surrounding remains unchanged after optical excitation. Our findings provide new insights into how fundamental electronic interactions function and demonstrate that dual optical excitation can transiently alter the coupling of electrons with lattice. This work not only advances our understanding of ultrafast magnetization dynamics but also suggests that the anomalous effects, such as increased relaxation times, observed in magnetically ordered materials may be a general phenomenon applicable to a wide range of systems. By uncovering the ability to control electron dynamics through dual optical excitation, this study opens new avenues for manipulating quantum states and designing next-generation ultrafast spintronic devices.

**Results**

**Experimental configuration. Figure 1a** shows the sketch of the experimental setup used in this study to measure the transient spin dynamics in Cu. Further details about the experimental configuration are given in the experimental section. Optical excitation was carried out with $\lambda$=1030 nm circularly polarized pump pulses. Pump 1 was incident on the sample at an angle of 2° and the pump 2 beam had an angle of 60°. The circular polarization was obtained with a zero-order quarter-wave plate. During the dual pump experiments, both pumps had the same chirality of circular polarization. The change in magnetization state was monitored with a much weaker, time-delayed $\lambda$=515 nm probe pulse, obtained by frequency doubling of the fundamental wavelength of the laser with a beta barium borate crystal. The incidence angle of the probe beam was 5° to the sample normal. The access to the time-resolved change in magnetization is realized via analysis of the rotation of the polarization of the probe pulse, making use of the magneto-optical Kerr effect in polar geometry and measured with a balanced photodiode. In this geometry, we measure the projection of the magnetic moment induced by pump pulses onto the wavevector of the probe light. Simultaneously, the time-resolved change in reflectivity was measured with a separate photodiode. All experiments were performed at room temperature and without an applied external magnetic field. The sample is a commercially available 5 x 5 x 1 mm one-sided polished single crystal of Cu with [110] being the out-of-plane direction.

**Inverse Faraday effect overview.** The IFE refers to the interaction between circularly polarized light and electrons in a material, where the light induces a change in the electron's spin state [24,25]. When circularly polarized light interacts with the electrons in a metal, it affects their orbital motion [26,27,28,29,30]. This alteration in orbital motion leads to an induce of orbital moment and, mediated by spin-orbit coupling, causes the spins of the electrons to flip. The result is an imbalance between the populations of spin-up and spin-down electrons, which leads to a transient change in the magnetic state of the material. However, it is important to note that this imbalance in the spin populations does not result in the formation of macroscopic magnetic order. Instead, the induced imbalance is a transient phenomenon, occurring on a sub-picosecond timescale, which does not lead to long-range spin alignment. As a result, the IFE provides a mechanism by which light can influence the electron spins on ultrafast timescales without the need for external magnetic fields [31,32]. The IFE has become a key phenomenon in the field of ultrafast magnetism, offering a tool for selective excitation of spin dynamics. Indeed, IFE has been used in experiments to study magnetization precession [33], to perform deterministic control of magnetization states [34], and is considered to be important for multi-pulse magnetization switching [35], making it highly relevant for applications in optomagnetic devices and spintronics [36].

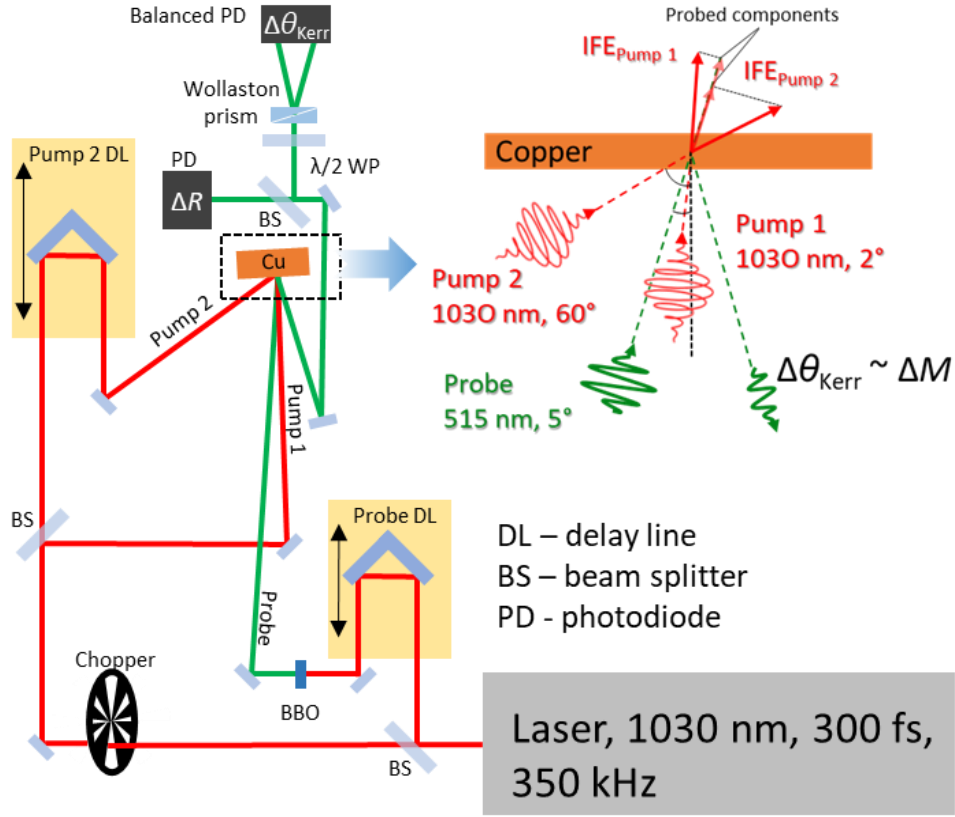

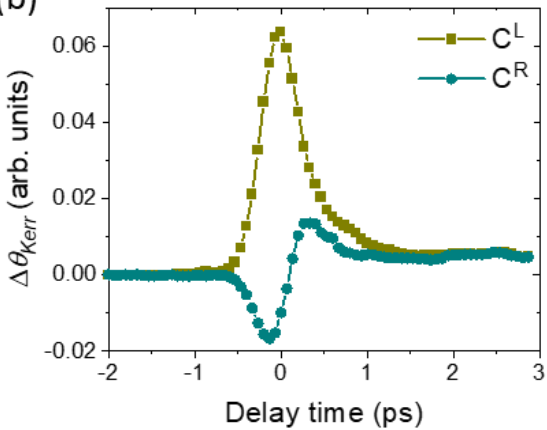
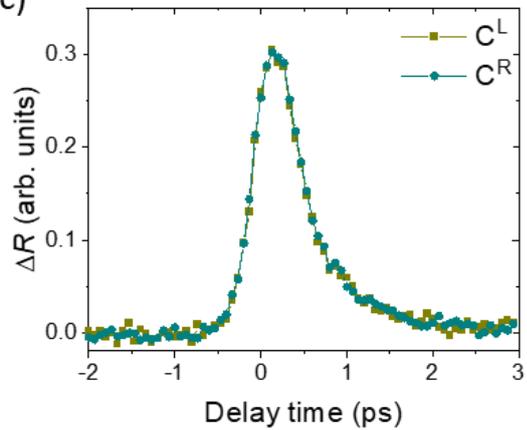

*Figure 1. (a) Experimental geometry for dual-pump time-resolved experiments. (b) and (c) Time-resolved differential polar Kerr signal and transient reflectivity of Cu, respectively, as a function of delay time after excitation with Pump 1 only with two different helicities of circular polarization and having a fluence of 16.9 mJ/cm$^2$.*

**Single pump dynamics.** In **Figure 1b**, the transient Kerr rotation signal, $\Delta\theta_{Kerr}$, is displayed for a single circularly polarized pump pulse (Pump 1) with two distinct helicities, measured at a fluence of 16.9 mJ/cm². The Kerr rotation signal is sensitive to changes in electron spin populations, reflecting the material's magnetization dynamics. The IFE alters these spin populations, inducing a transient change in the material's magnetic state. The differing traces arise from IFE-induced magnetization, where circularly polarized light of opposite helicities generates magnetization with opposing orientations. Ideally, this should produce peak-like Kerr rotation signals with identical amplitudes and shapes but opposite signs. However, the observed signals exhibit asymmetry, consistent with prior studies using similar experimental conditions [37]. This asymmetry does not stem from variations in excitation efficiency between the two helicities, as confirmed by the transient reflectivity data (**Figure 1c**), which show no significant differences between the two helicities. Instead, we attribute the asymmetry to a nonmagnetic contribution, detailed further in the Experimental Section.

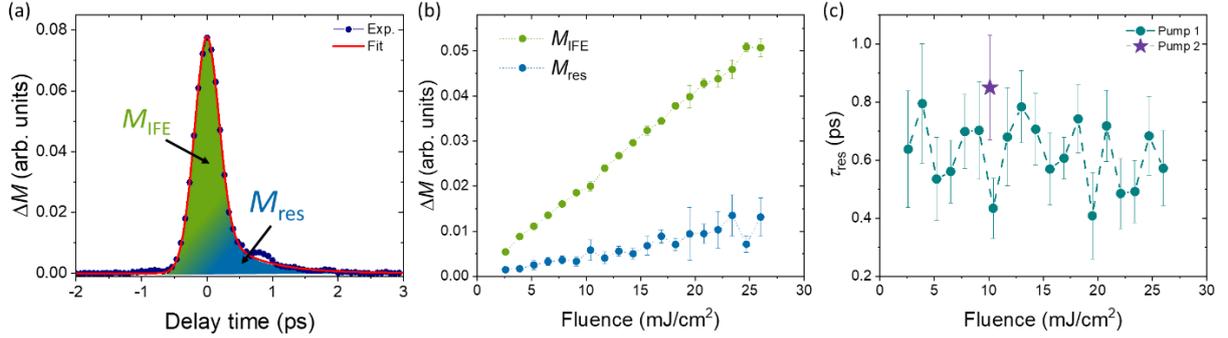

*Figure 2. (a) Light-induced magnetization as a function of time after excitation of Cu with Pump 1 only with a fluence of 16.9 mJ/cm². $M_{IFE}$ and $M_{res}$ components are indicated with green and blue, respectively. The solid red line is a fit to eq. (2). (b) Amplitudes of IFE $M_{IFE}$ and residual magnetization $M_{res}$, and (c) decay time of residual magnetization $\tau_{res}$ as a function of fluence. The error bars are a 95 % confidence interval.*

The magnetic contribution can be isolated by subtracting the time traces measured with the opposite helicity of circular polarization:

$$\Delta M(t) = \Delta\theta_{Kerr}^{C^L}(t) - \Delta\theta_{Kerr}^{C^R}(t), \qquad (1)$$

where $\Delta\theta_{Kerr}^{C^{L(R)}}(t)$ represents the time-resolved Kerr rotation signal measured when the pump beam is left(right)- handed circularly polarized. The resulting signal, shown in **Figure 2a**, was derived using the data from **Figure 1b** and depicts the spin dynamics after single-pulse excitation with Pump 1 at a fluence of 16.9 mJ/cm². The resulting magnetization signal reflects the imbalance between the number of spin-up and spin-down electrons, which is induced by the inverse Faraday effect (IFE). The data exhibits two key components: a peak-like signal at zero delay time, corresponding to the IFE-induced magnetization during the pulse excitation (shaded in green), and an additional exponential recovery of the residual spin imbalance (shaded in blue). The time-resolved magnetization traces were fitted with a combination of a Gaussian peak and exponential decay functions to capture both components:

$$\Delta M(t) = \Delta M_{IFE} e^{(t/\tau)^2} + \left\{\left(\Delta M_{res} e^{-t/\tau_{res}}\right) g(t)\right\} \otimes \Gamma(t) \qquad (2)$$

where $\Delta M_{IFE}$ and $\Delta M_{(res)}$ are the amplitudes of the IFE and residual magnetization components, $\tau_{res}$ is the decay time of the residual magnetization, g(t) is a step function, and $\Gamma$(t) represents the convolution of the fit function with the Gaussian laser pulse. **Figure 2b** shows the amplitude of the IFE signal, $\Delta M_{IFE}$ (green), and the amplitude of the residual magnetic moment, $\Delta M_{res}$ (blue), as a function of pump fluence. Both amplitudes increase linearly with increasing fluence, with a constant ratio of $\Delta M_{IFE} / \Delta M_{res} = 3.5$. This linear increase is as expected since the IFE is an opto-magnetic effect, i.e., $\Delta M \propto E^2 \propto$ Fluence, where $E$ is the electric field strength. As pump fluence rises, the number of photon-induced spin-flip events increases, enhancing the spin imbalance and increasing both the $\Delta M_{IFE}$ and $\Delta M_{res}$ signals proportionally. Within the fluence regime during our experiments, we do not reach the nonlinear excitation regime [38]. The characteristic decay time of the residual spin imbalance, $\tau_{res}$, quantifies the rate of angular momentum transfer from electrons to the lattice following optical excitation. **Figure 2c** shows $\tau_{res}$ as a function of pump fluence for single-pump excitation. Across all fluences, $\tau_{res}$ remains at approximately 0.6 ps, consistent with expectations for single-pump excitation [39]. This stability indicates that the angular momentum transfer efficiency is unaffected by light intensity. Notably, $\tau_{res}$ reflects the ability of individual electrons to transfer angular momentum to the lattice, independent of macroscopic magnetic order.

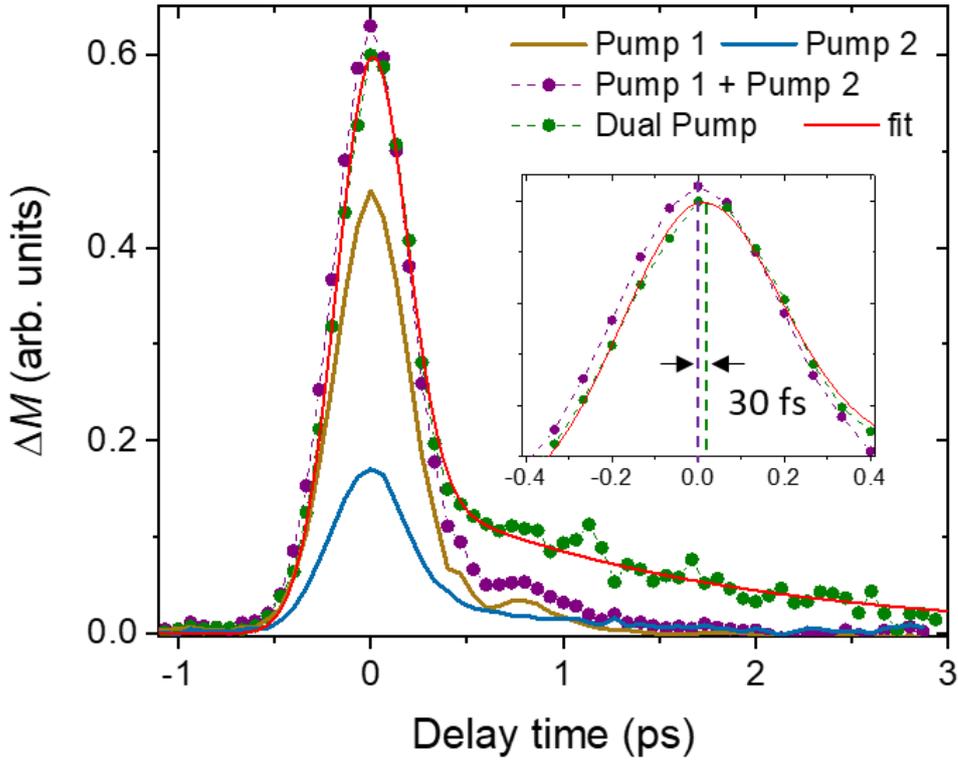

*Figure 3. Transient magnetic signal $\Delta M$ as a function of delay time after dual optical excitation (green), shown together with traces for individual Pump 1 (dark yellow) and Pump 2 (blue) excitation, as well as the linear superposition of signal from Pump 1 and Pump 2 (purple). The laser fluence of both pump pulses was 10.1 mJ/cm$^2$. The solid red line is a fit to eq. (2) Inset: enlarged view of the $\Delta M_{IFE}$ component.*

**Dual pump dynamics.** In the dual optical excitation experiments, both Pump 1 and Pump 2 were circularly polarized with the same helicity, and the magnetic signal was extracted in the same manner as for single-pump excitation. The fluence for both pump beams was 10.1 mJ/cm². **Figure 3** shows the transient magnetic signal $\Delta M$ as a function of delay time after excitation with Pump 1 only (yellow solid line) and Pump 2 only (blue solid line). The signal from the Pump 2 excitation closely resembles the dynamics induced by Pump 1, although the amplitude of $\Delta M$ for Pump 2 excitation is about half that of Pump 1. This reduction in amplitude is due to the experimental geometry, where the projection component of $\Delta M$ for Pump 2 is measured at an angle of 55° to the probe beam (see **Figure 1a**). The decay time for the Pump 2 excitation was found to be $\tau_{res} = 0.85 \pm 0.18$ ps (see **Figure 2c**), which is in good agreement with the decay time for Pump 1 excitation (see **Figure 2c**). This confirms that the individual effects of both pump beams on electron dynamics are similar. However, when both pumps were applied simultaneously (dual-pump excitation), the transient magnetic response, shown in green, was noticeably different. The decay time increased to $\tau_{res} = 1.54 \pm 0.22$ ps, which is approximately 2.5 times larger than the decay time observed for single-pump excitation. Furthermore, when applying the same fitting procedure for dual-pump excitation, the obtained amplitude of the IFE-induced signal, $\Delta M_{IFE}$ (shown in green), was found to be about 3% smaller than the expected amplitude based on the linear superposition of the individual pump signals shown in purple, and the ratio $\Delta M_{IFE} / \Delta M_{res} = 3.1$ is smaller compared to single-pump experiments. Additionally, a shift of about 30 fs in the $\Delta M_{IFE}$ peak position relative to the linear superposition trace was observed (see the inset to **Figure 3a**).

**Discussion**

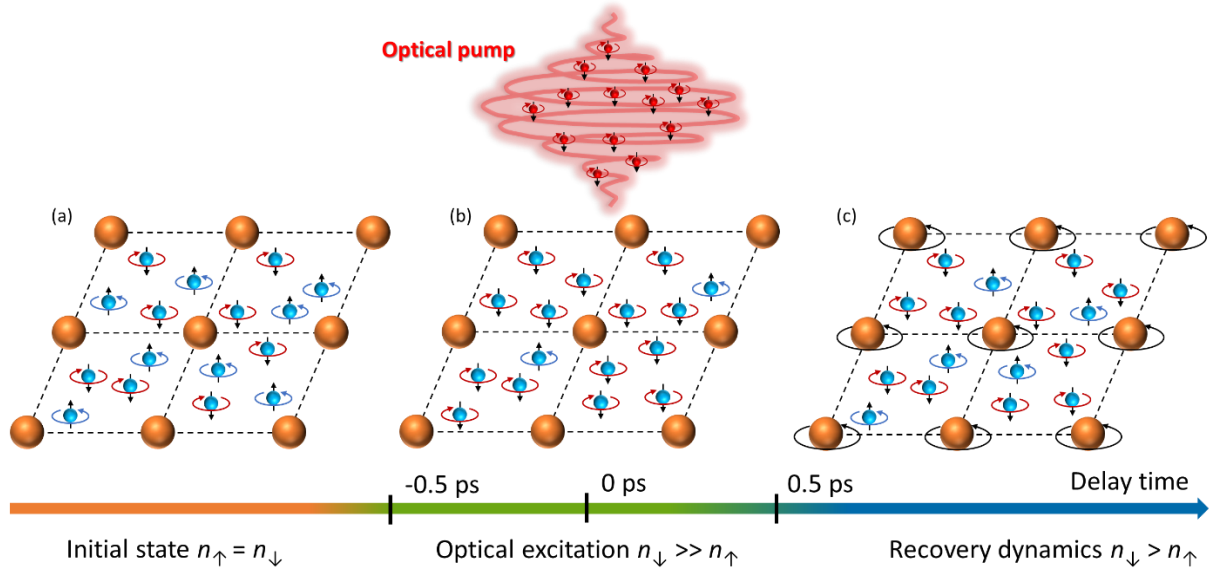

*Figure 4. Schematic representation of ultrafast spin dynamics in copper. Although the optical excitation is illustrated as a single circularly polarized pulse, the schematic represents dynamic steps that apply equally to both single- and dual-pump excitation. Panel (a) depicts the equilibrium state of copper, where equal numbers of spin-up and spin-down electrons result in no macroscopic magnetic moment. The spin of electrons (blue spheres) is indicated with black*

*arrows and associated angular momentum with circular arrows in blue and red for opposite spin direction. The Cu ions are depicted with orange spheres. Panel (b) illustrates the state during optical excitation, where circularly polarized photons (red spheres) induce a transient spin imbalance via the inverse Faraday effect—here, a small fraction of electrons flip their spin (the schematic exaggerates this for clarity). Panel (c) shows the recovery phase after the pump pulse, as the residual spin imbalance decays due to angular momentum transfer from the electrons to the lattice via the generation of polarized phonons.*

The observed difference between single- and dual-pump experiments is striking and challenges our fundamental understanding of spin-polarized electron dynamics in metals after optical excitation. To gain a more intuitive understanding of these results, one can consider **Figure 4**, which schematically shows the different phases of the dynamical process. In equilibrium, copper exhibits no macroscopic magnetic moment, with an equal number of spin-up and spin-down electrons (**Figure 4a**). Upon optical excitation, a small fraction of electrons undergo spin flips, creating a transient magnetic moment, $M_{IFE}$, as shown in **Figure 4b**. For clarity, the number of spin flips in the schematic is exaggerated; in reality, only about 1 in 1000 optically excited electrons flip their spin. Most of these spins relax immediately after the laser pulse ends, with a small fraction retaining the angular momentum transferred from light during illumination. This residual magnetic moment $M_{res}$ decays over time as electrons transfer angular momentum to the lattice by generating phonons with angular momentum, typically occurring on a sub-picosecond timescale [40,41] (**Figure 4c**). The contradiction to our current understanding of electron dynamics in metals after optical excitation arises when considering the dynamics after the laser pulse ends. Whether the spins were flipped by a single pump or dual pumps, the electrons should, in principle, behave identically once the excitation is over. After approximately 0.5 ps, the electrons are no longer influenced by the pump light or any macroscopic magnetic arrangement and interact freely with their environment. Furthermore, the periodic excitation pattern [42,43,44,45,46,47] caused by the interference of the two pump beams should not affect the electron dynamics, as evidenced by the nearly identical decay times observed across different fluences in single-pump experiments (**Figure 2c**). Yet, the electrons excited by dual pumps exhibit significantly different dynamics, with a 2.5-fold increase in the decay time $\tau_{res}$. This implies that dual optical excitation has an additional action that modifies the efficiency of angular momentum exchange between electrons and the lattice. Since this occurs when the excitation densities of the dual and single pumps are the same, it challenges the conventional view that optical excitation merely increases the electron energy.

**Mechanistic Insights.** The anomalous increase in the decay time, $\tau_{res}$, of the residual spin imbalance following dual-pump excitation suggests that electrons after dual optical excitation transfer angular momentum to the lattice less efficiently than those excited by a single pulse. One plausible explanation is that dual optical excitation induces a transient, long-lasting modification of the electron wavefunction, which alters the dynamics of electrons. This modification persists significantly beyond the duration of the pump pulses, as demonstrated by the delay time range where discrepancies between single-pulse and dual-pulse dynamics are observed, and causes a reduced rate of angular momentum transfer from electrons to the lattice. The data presented here for Cu are consistent with previous experimental observations of spin dynamics in magnetically ordered materials, where dual optical excitation similarly reduced the efficiency of angular momentum transfer in various systems. For example, experiments on ultrafast demagnetization dynamics in Pt/Co/Pt and TbCo systems [21] revealed increased recovery times following ultrafast suppression of magnetization induced by dual-pump excitation compared to single-pulse excitation, occurring on picosecond timescales. Similarly, studies of magnetization precession in permalloy [23] showed prolonged decay times for magnetic oscillations triggered by dual-pump excitation compared to single-pump excitation, persisting for hundreds of picoseconds – far longer than any known light-induced modifications to electronic states in metals. These consistent trends observed across different materials and timescales suggest that the phenomenon may represent a general effect.

**Broader Implications.** The ability to modify the spin dynamics at the level of individual electrons through dual optical excitation opens exciting possibilities for controlling quantum states in materials. This technique could be harnessed to manipulate spin dynamics in ultrafast spintronic devices, where precise control over angular momentum transfer is essential. Moreover, the persistence of modified electron dynamics over nanosecond timescales suggests that dual optical excitation could enable long-lasting changes in material properties without altering the crystal structure. This presents a novel, non-thermal route to functionalizing materials properties for applications in optomagnetic devices and materials for quantum technologies.

From a fundamental perspective, our results challenge the conventional assumption that optical excitation can only change the occupation of electronic energy levels without modifying the intrinsic properties of individual electrons after the excitation and subsequent relaxation processes. The observation that dual optical excitation can alter the efficiency of angular momentum transfer suggests that light can influence certain intrinsic properties of electrons in specific ways. This discovery may have profound implications for our understanding of electronic interactions in solids and could pave the way for new theoretical models to account for such effects.

**Future Directions.** While this study provides compelling evidence for the modification of electronic dynamics by dual optical excitation, there are still many questions to be addressed. For instance, is this effect specific to metals, or can it be observed in other nonmagnetic or magnetic materials? How do factors such as excitation wavelength and pulse duration influence the observed dynamics? Addressing these questions will require systematic experimental investigations across a range of materials and conditions, as well as theoretical work based on significantly more experimental data to develop models that explain the underlying mechanisms.

**Experimental Section**

*Sample Preparation and Characterization.* The sample used in this study was a commercial single crystal of copper (5 × 5 × 1 mm, [110] out-of-plane orientation) purchased from MTI Corporation. One side of the sample was polished to ensure a smooth, reflective surface suitable for laser excitation and measurements. No additional surface treatments were applied prior to the experiments.

*Laser System and Measurement Techniques.* Ultrashort laser pulses were generated using an ActiveFiber laser system, delivering pulses with a central wavelength of 1030 nm, a duration of about 300 femtoseconds, and a repetition rate of 350 kHz. The fundamental wavelength for the pump ($\lambda$=1030 nm) was chosen for its stability and avoids complications associated with frequency conversion. To produce a $\lambda$=515 nm wavelength probe pulse for time-resolved measurements, we employed frequency doubling using a beta barium borate crystal. The dual-pump excitation setup utilized two laser pulses (Pump 1 and Pump 2) incident on the sample at angles of 2° and 60°, respectively. These angles were chosen to replicate the experimental geometry used in prior studies on magnetically ordered materials, ensuring consistency with earlier experiments. The beam radius was 250 µm for the pump and 100 µm for the probe at normal incidence. Circular polarization for pump pulses was obtained with a zero-order quarter-wave plate. The pump beam intensity was modulated with an optical chopper at 500 Hz. Measurements were conducted at fluences 35% below the single-pump damage threshold of 35 mJ/cm² to ensure non-destructive conditions.

Magnetization dynamics was measured using a balanced photodiode setup to detect the transient polar Kerr rotation caused by the induced magnetization. The probe beam was split into vertically and horizontally polarized components using a Wollaston prism, and the difference in signal between the two components was measured to determine the polarization rotation of the probe beam. This rotation is directly sensitive to the magnetization of the sample. A narrow-band filter (520 nm ± 40 nm) was placed before the detector to ensure that only the 515 nm probe light was measured, eliminating any contamination from pump light or other wavelengths. However, when the sample is excited, there is also a change in reflectivity, which leads to a change in the overall intensity of the reflected probe beam. This change affects the difference signal on the balanced photodiode and causes an asymmetric shape of Kerr rotation signal for opposite helicity of the pump beam (see **Figure 1b**) but does not alter the interpretation of the data because we take the difference between traces taken with opposite helicity of the pump pulse to extract the magnetic signal. In addition to Kerr rotation measurements, transient reflectivity changes were monitored using a separate photodiode. Both signals were collected simultaneously using two lock-in amplifiers, enabling a precise correlation between magnetization dynamics and reflectivity changes under identical experimental conditions.

**Acknowledgments:**

The authors express their sincere gratitude to Laura Heyderman for her valuable assistance in the final editing of the manuscript.